\documentclass[preprint,showpacs,preprintnumbers,amsmath,amssymb]{revtex4}

\usepackage{graphicx }
\usepackage{dcolumn}
\usepackage{bm}
\usepackage{amsmath}

\begin{document}


\title{Aperiodicity in one-way Markov cycles and repeat times of large earthquakes in faults}

\author{Alejandro Tejedor}
 \email{atejedor@unizar.es}
 \affiliation{Theoretical Physics Department, University of Zaragoza.}
\author{Javier B. G\'omez}
 \email{jgomez@unizar.es}
\affiliation{
Earth Sciences Department, University of Zaragoza.
}%
\author{Amalio F. Pacheco}%
 \email{amalio@unizar.es}
\affiliation{%
Theoretical Physics Department, University of Zaragoza, and BIFI
}%

\date{\today}

\begin{abstract}
A common use of Markov Chains is the simulation of the seismic cycle in a fault, i.e. as a renewal model for the repetition of its characteristic earthquakes. This representation is consistent with Reid's elastic rebound theory. Here it is proved that in {\it any} one-way Markov cycle, the aperiodicity of the corresponding distribution of cycle lengths is always lower than one. This fact concurs with observations of large earthquakes in faults all over the world.  
\end{abstract}

\maketitle

\section{\label{sec:intro}Introduction}

The elastic-rebound model is the canonical ``macroscopic'' theory of great earthquakes \cite{Reid10,Scholz02}. It states that a great earthquake will occur where large elastic strains have accumulated in the crust. The earthquake itself will relieve most of the strain which will then accumulate slowly again by a steady input of tectonic stress until the elastic strain becomes sufficiently large for another earthquake to ensue. The duration of this ``earthquake cycle'' is the ratio of the strain released during an earthquake to the rate of input of tectonic strain by plate motion.

A corollary of the elastic-rebound model is the concept of the characteristic earthquake \cite{Schwartz84}. Although a specific seismic fault or fault segment can produce small earthquakes as well as large ones, an overwhelming part of the stored elastic energy is released by the large ones, which tend to rupture the entire area of the fault (or fault segment). As the magnitude of an earthquake is related to the broken area of the fault \cite{Kanamori75}, each fault (or fault segment) tends to produce large earthquakes of the same magnitude; and because these earthquakes release most of the stored elastic energy, their repetition defines the duration of the earthquake cycle.

Because the Earth's crust is heterogeneous and faults are not isolated from each other but communicate through long-range stress-transfer mechanisms \cite{Stein05}, the earthquake cycle is not periodic. So, although the elastic-rebound model is in essence deterministic, its application to a heterogeneous and interacting crust implies its translation into a probabilistic framework. Only in this way can it be used for earthquake forecasting purposes.  

Several authors have proposed probabilistic versions of the elastic-rebound model, in the shape of probability distribution functions (pdfs) for the duration of the earthquake cycle \cite{Rikitake74, Utsu84, Matthews02, Vazquez02, Michael05, Gonzalez06a}. The rationale of these pdfs ranges from purely statistical (e.g. \citet{Utsu84}) to physically-motivated (e.g.\citet{Vazquez02}). However, due to the scarcity of registered large earthquakes in a specific fault (usually 4 to 10 earthquakes), the statistics upon which the selection of a specific pdf is based are poor. This means that different pdfs can fit the empirical distribution function. In any case, any pdf to be used for these purposes should have at least the following properties:

\begin{enumerate}
\item As the time between successive earthquakes is a positive quantity, the substrate of the pdf must be the positive part of the real line.
\item Very short cycles, as compared with the mean of the distribution, are extremely rare due to the physics of the elastic-rebound model. This means that the potential pdfs must be zero (or nearly so) for small values of its argument. This period of zero (or very small) probability at the beginning of the earthquake cycle is called stress shadow in seismology. 
\item The probability of cycles much longer than the mean duration is a decreasing function of the duration, and tends to zero as the duration tends to infinity.

\end{enumerate}

Although the complete pdf gives more information than any of its moments or other statistical parameters derived from it, summarizing the information content of the pdf in one or a few numbers is very convenient, more so when the pdf does not have a close analytical form or is difficult to obtain. Focusing attention on one of their statistical parameters benefits the comparison of the performance of different pdfs. One of these parameters is the coefficient of variation, $\alpha$, the ratio of the standard deviation to the mean of the pdf. 

In the seismological literature the coefficient of variation is known as the aperiodicity, a very descriptive name when applied to the duration of the earthquake cycle: when  $\alpha=0$ the earthquake cycle is perfectly periodic, when $0 < \alpha  < 1$ the earthquake cycle is quasiperiodic, and when   $\alpha> 1$ the earthquake cycle is said to have a {\it clustering} of events. The case $ \alpha=1 $ is particularly important because the exponential distribution has this property, and the exponential distribution is the pdf of an earthquake cycle where large earthquakes occur in time following a Poisson distribution (i.e, they are random in time).

The predictability of a time series whose events follow a specific pdf is related to its aperiodicity \citep{Tejedor09a}. Applied to the earthquake cycle this means that the predictability of the next large (characteristic) earthquake in a series is related to the aperiodicity of the pdf describing the duration of the cycles: aperiodicities close to zero imply greater predictability than aperiodicities close to one.  \citet{Sykes06} have calculated the aperiodicity of the earthquake cycle of several seismic faults. All the studied faults have aperiodicities smaller than $0.6$, meaning that the earthquake cycle is quasiperiodic.  \citet{Ellsworth99} also studied the aperiodicity of the earthquake cycle in several fault segments and concluded that all of them are between $0.11$ and $0.97$. It seems that  $\alpha < 1$ is a property of the earthquake cycle in seismic faults. Can this be reproduced by simple models of single-fault seismicity?

Using the technique of Markov chains \cite{Durrett99}, \citet{Vazquez02} built a simple cellular automaton model (the minimalist model) capable of reproducing the main characteristics of the seismicity of individual faults. This same model was later used as a renewal model of seismicity \cite{Utsu84, Vere70}  and applied to the forecasting of the next earthquake in the Parkfield, California, segment of the San Andreas Fault \cite{Gonzalez06a}. The aperiodicity of the earthquake cycle in the minimalist model is always smaller than $0.5$.

\citet{Gonzalez05} proposed another physically-motivated Markov model of the earthquake cycle (the box model). It also has characteristic-earthquake behaviour but, unlike the minimalist model, it only produces earthquakes of one size, that of the maximum possible magnitude given the area of the fault. This simplification  is acceptable as large earthquakes release most of the stored energy. The aperiodicity of the earthquake cycle in this model is also lower than 1.  

A question that we want to answer in this paper is whether a general {\it one-way} Markovian model of the earthquake cycle can be constructed with aperiodicities larger that 1.

\begin{figure}
\noindent \includegraphics[width=16pc]{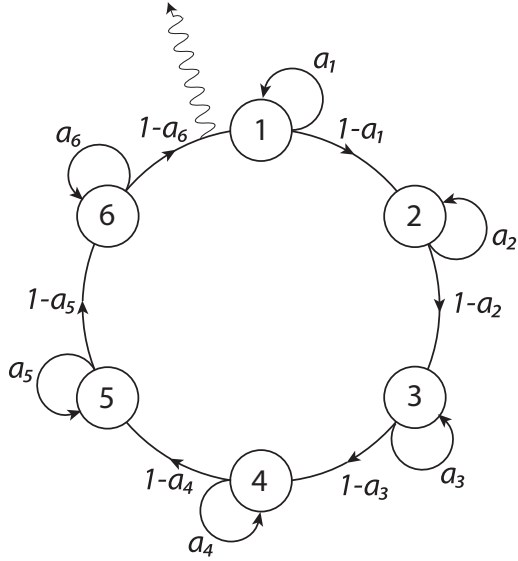}
\caption{\label{One_way_figure} Scheme of a one-way Markov cycle with $N=6$}
\end{figure}

The term one-way in this family of models refers to the fact that after a time step, the state of strain can remain either stationary or grow by a finite amount. In other words, in this type of model a decrease in the strain, such as could take place in a random walk model, is forbidden. Time increases in discrete steps and strain is  also added in finite units. The $N$ positions of the model correspond to states of the system with progressive growing strain.  The scheme of this model is shown in figure \ref{One_way_figure}, for $N=6$.

The relaxation of the system through a sudden and complete loss of strain, which simulates the occurrence of an earthquake, occurs when the $N^{th}$ position of chain is reached. In fig. \ref{One_way_figure}, the relaxation is represented by the wavy line.

This article is organized as follows: Section \ref{OneWayMarkov} contains the general form of the stochastic matrix of one-way Markov cycles together with the specialization to the case of the Box-Model and the case where all the parameters are equal. Sections \ref{Distribution_functions} and \ref{Moments} contain the distribution function for the Cycle Length and the two first moments of that distribution, respectively. The distribution function and first moments of the two particular cases mentioned above are also included.
Adducing that a One-Way Markov Cycle is a succession of $N$ independent geometric processes, a simplified form of the mean and variance are written in Section \ref{Simply_moments}.   In Section \ref{Forecasting}, the so called fraction of error and fraction of alarm time are calculated.  In Section \ref{Inequality_aperiodicity}, we prove that in any one-way Markov cycle, the aperiodicity of the corresponding distribution of cycle lengths is always lower than one. Finally, in Section \ref{Conclusions} we write the conclusions. 
Additionally, we have considered it of interest  to explicitly present, for a non trivial case such as $N=3$, how the distribution function in the case where all the parameters are equal tends to the Negative Binomial Distribution. This proof is written in the Appendix.

\section{One-way Markov cycles.  Two particular cases} \label{OneWayMarkov}

The Markov matrix $[M]$ of a  one-way Markov cycle of size $N$ is like this:

\[
[M] =
 \begin{pmatrix} \label{Matrix_Markov}
  a_1 & 1-a_1 & 0 &  \cdots & 0 & 0 & 0 \\
 0 & a_2 & 1-a_2 & \cdots & 0 & 0 & 0  \\
 0 & 0 & a_3  & \cdots & 0 & 0 & 0  \\
  \vdots  & \vdots  & \vdots & \ddots & \vdots & \vdots & \vdots \\
 0 & 0 & 0 & \cdots & a_{N-2} & 1-a_{N-2} & 0  \\
 0 & 0 & 0  & \cdots & 0 & a_{N-1} & 1-a_{N-1}  \\
 1-a_N & 0 & 0  & \cdots & 0 & 0 & a_N  \\
 \end{pmatrix}
\]
\\

\noindent where the $N$ parameters $a_i$, $i=1,2,3,..,,N$, are $0<a_i<1$.  
There are $N$ positions in the cycle. In each basic step, the system, with probability $a_i$, remains in the same position, and with probability $(1-a_i)$  jumps to the next one. This system is illustrated in Fig. \ref{One_way_figure}.
Denoting by $[M]^T$  the transpose of the Markov Matrix, the components of its eigenvector with eigenvalue unity, $c_i$ , are:

\begin{eqnarray} \label{eingenvalue}
c_i=\frac{1}{C}\prod^{N}_{j(\neq i)=1}(1-a_j)  \nonumber \\
\\
C=\sum^{N}_{i=1}\prod^{N}_{j(\neq i)=1}(1-a_j) \nonumber  \\    \nonumber
\end{eqnarray}

\noindent where $C$ is the normalization factor. 
The component $c_i$ is the probability, statistically speaking, of finding the system in the position i of the cycle. The value of the N components expressed in \ref{eingenvalue} are easily obtained from systematics.

A particular case of this general scheme is that of the Box Model (BM) \cite{Gonzalez05}. As said in the introduction, this is a cellular automaton where the stochastic filling of a box represents the loading of elastic energy in a seismic fault. The emptying of the box after it is full is analogous to the generation of an earthquake. In this model, the value of the $N$ parameters is:

\begin{equation} \label{BoxM}
a_i=\frac{i-1}{N}
\end{equation}

Another significant particular case corresponds to the case when 

\begin{equation}  \label{EqualProb}
a_i=a  \qquad  \forall i
\end{equation}
                                      
That is, all the parameters are identical. In such a Markov process, the distribution of the cycle length is that of a Negative Binomial Distribution (NBD) \cite{Canavos84} where the probability of success is $(1-a)$ and $N$ successes are required.

\section{Distribution Function for the cycle lengths}\label{Distribution_functions}

Eliminating the element $(N,1)$ of the Markov matrix in matrix $[M]$, i. e. substituting $(1-a_N)$ by 0, we obtain a new ``pruned matrix'' $[M]'$  which provides the distribution of the cycle lengths, $P_N(n)$. Letter $n$ represents the length of the cycle expressed in time steps of the model. Thus, $n$ is a positive integer and $P_N(n)$ is the probability that in a system of size $N$, the cycle is completed in $n$ steps. The formula for this probability is

\begin{equation}  \label{Prob_back}
P_N(n)=[M]^{n-1}_{1,N}·(1-a_N)
\end{equation}

\noindent which expresses that the probability that the length cycle is $n$ is equal to the product of two factors. The first  is the probability that having departed from stage 1, after $n-1$ steps the system is in stage $N$.  The second is the probability of passing in one step from position $N$ to position 1. In the first factor, the use of $[M]'$  guarantees that in the first $n-1$ steps there has been no transition between these two positions.
The computation of $P_N(n)$ provides:

\begin{eqnarray} \label{PN_de_n}
P_N(n)=[\prod_{i=1}^{N}(1-a_i)][\sum_{i=1}^{N}\frac{a_{i}^{n-1}}{\prod^{N}_{j(\neq i)=1}(a_i-a_j)}] \nonumber \\
\\
n = N, N+1, \dots, \infty \nonumber  \\   \nonumber
\end{eqnarray}

This is the general form of the discrete distribution function in any one-way Markov cycle. This formula has been also obtained from systematics.

As stated in the Introduction, when these systems are applied in seismicity, the fact that until $n=N$ the probability of completing a cycle is null is called stress shadow.

In the two particular cases mentioned above one obtains:

(BM)

\begin{eqnarray}\label{PN_de_n_BOX}
P_N(n)=\sum_{i=1}^{N-1}(-1)^{i+1}\binom{N-1}{i-1}(1-\frac{i}{N})^{n-1} \nonumber \\
\\
n = N, N+1, \dots, \infty \nonumber  \\   \nonumber
\end{eqnarray}

\noindent and (NBD)

\begin{eqnarray} \label{PN_de_n_BinNeg}
P_N(n)=(1-a)^{N}a^{n-N}\binom{n-1}{N-1} \nonumber \\
\\
n = N, N+1, \dots, \infty  \nonumber \\   \nonumber
\end{eqnarray}

\section{The two first moments}\label{Moments}

Using eq. \ref{PN_de_n} the mean length of the cycles, $\mu$, is given by:

\begin{equation}  \label{mean}
\mu=\langle n \rangle = [\prod_{i=1}^{N}(1-a_i)][\sum_{i=1}^{N}\frac{1}{(1-a_i)^2\prod^{N}_{j(\neq i)=1}(a_i-a_j)}]
\end{equation}

 And the variance, $\sigma^2$, is given by

\begin{equation}  \label{sigma}
\sigma^2=\langle n^2 \rangle - \mu ^2= [\prod_{i=1}^{N}(1-a_i)][\sum_{i=1}^{N}\frac{1-a_i^2}{(1-a_i)^4\prod^{N}_{j(\neq i)=1}(a_i-a_j)}]-\mu^2
\end{equation}

In the BM

\begin{equation}  \label{mean_BM}
\mu=1+\sum_{i=2}^{N}\frac{N}{N+1-i}
\end{equation}

\noindent and 

\begin{equation}  \label{sigma_BM}
\sigma^2=\sum_{i=2}^{N}\frac{1-\frac{N+1-i}{N}}{(\frac{N+1-i}{N})^2}
\end{equation}

In the NBD

\begin{equation}  \label{mean_NBD}
\mu=\frac{N}{1-a}
\end{equation}

\noindent and

\begin{equation}  \label{sigma_NBD}
\sigma^2=\frac{aN}{(1-a)^2}
\end{equation}

\section{Simplifying the two first moments}\label{Simply_moments}

In a geometric process where the probability of success is $(1-a)$, the mean and variance of the distribution are:

\begin{eqnarray} \label{geometric}
\mu=\frac{1}{1-a} \nonumber \\
\\
\sigma^2=\frac{a}{(1-a)^2} \nonumber  \\   \nonumber
\end{eqnarray}

Then, as a One-Way Markov Cycle is nothing more than a succession of $N$ independent geometric processes, the mean and variance can be written as:

\begin{equation}  \label{mean_Markov}
\mu=\frac{1}{1-a_1}+\frac{1}{1-a_2}+\dots+\frac{1}{1-a_N}
\end{equation}

\noindent and 

\begin{equation}  \label{sigma_Markov}
\sigma^2=\frac{a_1}{(1-a_1)^2}+\frac{a_2}{(1-a_2)^2}+\dots+\frac{a_N}{(1-a_N)^2}
\end{equation}

The reader will note the marked difference between  Eqs. \ref{mean} and \ref{sigma}, which were derived directly from Eq. \ref{PN_de_n} and Eqs. \ref{mean_Markov} and \ref{sigma_Markov} obtained by invoking the simplifying argument.
Thus, the aperiodicity, $\alpha$, is given by:

\begin{equation}  \label{aperiodicity_Markov}
\alpha=\frac{[\frac{a_1}{(1-a_1)^2}+\frac{a_2}{(1-a_2)^2}+\dots+\frac{a_N}{(1-a_N)^2}]^{1/2}}{\frac{1}{1-a_1}+\frac{1}{1-a_2}+\dots+\frac{1}{1-a_N}}
\end{equation}

\section{Fraction of error and fraction of alarm}\label{Forecasting}
A simple earthquake forecasting strategy consists of turning on an alarm at a fixed value of $n$ time steps after the last earthquake and maintaining the alarm until the next one.  Then the alarm is switched off and the same strategy is repeated. \cite{Newman02,Vazquez03,Keilis03}

For any thinkable strategy based on the use of alarms, if an earthquake takes place when the alarm is on, the prediction is considered to be a success. If it takes place when the alarm is off, there has been a failure to predict.
Then, denoting by fraction of error, $f_e$, the number of prediction failures divided by the total number of earthquakes, and the fraction of alarm, $f_a$, as the ratio of the time during which the alarm is on to the total time of observation, in our model and using the above mentioned strategy these two functions adopt the form:

\begin{equation} \label{Error_fraction_1}
f_e(n)=\sum_{n'=N}^{n}P(n')=1-\sum_{n'=n+1}^{\infty}P(n')
\end{equation}

Note that $f_e$ is the accumulated distribution of eq. \ref{PN_de_n}. Performing the sum in the $n'$ index, the result is:

\begin{equation}\label{Error_fraction_2}
f_e(n)=1-[\prod_{i=1}^{N}(1-a_i)]\sum_{i=1}^{N}[\frac{a_i^n}{(1-a_i)}\frac{1}{\prod^{N}_{j(\neq i)=1}(a_i-a_j)}]
\end{equation}

Regarding $f_a$,
\begin{equation} \label{Alarm_fraction_1}
f_a(n)=\frac{\sum_{n'=n}^{\infty}P(n')(n'-n)}{\sum_{n'=n}^{\infty}P(n')n'}=\frac{\sum_{n'=n}^{\infty}P(n')(n'-n)}{\mu}
\end{equation}

\noindent the result is:
\begin{eqnarray} \label{Alarm_fraction_2}
\mu·f_a(n)=\sum_{n'=n}^{\infty}n'P(n')-n\sum_{n'=n}^{\infty}P(n')=   \\ \nonumber
=[\prod_{i=1}^{N}(1-a_i)]\sum_{i=1}^{N}[\frac{a_{n}^{i}}{\prod_{j(\neq i)=1}^{N}(a_i-a_j)}·(\frac{n}{a_i(1-a_i)}+\frac{1}{(1-a_i)^2})]-n(1+P(n)-f_e(n))
\end{eqnarray}

\section{Inequality in the aperiodicity}\label{Inequality_aperiodicity}

Knowing the value of the maximum and minimum parameters in the system, $a_M$ and $a_m$, from eq. \ref{aperiodicity_Markov} the aperiodicity is bounded by the following inequality:

\begin{equation}  \label{aperiodicity_bounding1}
0 < \sqrt{\frac{a_m}{N}}(\frac{1-a_M}{1-a_m}) \leq \alpha \leq \sqrt{\frac{a_M}{N}}(\frac{1-a_m}{1-a_M})
\end{equation}

If all the parameters are equal to $a$, eq. \ref{aperiodicity_bounding1} shrinks to

\begin{equation}  \label{aperiodicity_bounding2}
0< \sqrt{\frac{a}{N}} \leq \alpha \leq \sqrt{\frac{a}{N}}
\end{equation}

\noindent or equivalently

\begin{equation}  \label{aperiodicity_NBD}
\alpha = \sqrt{\frac{a}{N}}
\end{equation}

\noindent which is the aperiodicity of the NBD; see eqs \ref{mean_NBD} and \ref{sigma_NBD}.

Looking for a general inequality in $\alpha$, from eq. \ref{aperiodicity_Markov} we deduce

\begin{eqnarray} \label{aperiodicity_bounding3}
\alpha = \frac{[\frac{a_1}{(1-a_1)^2}+\frac{a_2}{(1-a_2)^2}+\dots+\frac{a_N}{(1-a_N)^2}]^{1/2}}{\frac{1}{1-a_1}+\frac{1}{1-a_2}+\dots+\frac{1}{1-a_N}} \leq \nonumber \\ 
\\
\leq \frac{\frac{a_1^{1/2}}{1-a_1}+\frac{a_2^{1/2}}{1-a_2}+\dots+\frac{a_N^{1/2}}{1-a_N}}{\frac{1}{1-a_1}+\frac{1}{1-a_2}+\dots+\frac{1}{1-a_N}} \leq 1  \nonumber \\ \nonumber
\end{eqnarray}

The first inequality is obvious by comparing the square of the two numerators:

\begin{eqnarray}  \label{aperiodicity_bounding4}
\frac{a_1}{(1-a_1)^2}+\frac{a_2}{(1-a_2)^2}+\dots+\frac{a_N}{(1-a_N)^2} \leq \nonumber \\ 
\\
\leq \frac{a_1}{(1-a_1)^2}+\frac{a_2}{(1-a_2)^2}+\dots+\frac{a_N}{(1-a_N)^2}+\frac{1}{2}\sum_{i,j=1}^{N}\frac{a_i^{1/2}}{(1-a_i)}\frac{a_j^{1/2}}{(1-a_j)} \nonumber \\ \nonumber
\end{eqnarray}

This is because all the $a_i$ are positive and lower than 1.

The second inequality in eq. \ref{aperiodicity_bounding3} is also obvious because each of the $N$ terms in the numerator is smaller than its corresponding term in the denominator.

\section{Conclusions}\label{Conclusions}

We have calculated the form of the distribution function for the cycle length of any finite one-way Markov cycle. The number of independent parameters, $N$, coincides with the number of positions in the Markov cycle.  The first moments of this distribution are easily calculated bearing in mind that a one-way Markov cycle is nothing more than a succession of $N$ independent geometric processes. Thus, these moments are written as the sum of the mean, or variance, of the $N$ stages of the cycle.

The above enumerated properties of the model nicely correspond to  Reid's theoretical vision of the mechanism of how earthquakes are generated. As commented on in the Introduction, data of the time of recurrence of main shocks in faults all over the world indicate that the aperiodicity of those data is always lower than unity. This phenomenological fact agrees with the rigorous upper bound established here: the aperiodicity of the distribution of cycle lengths in any one-way Markov Cycle is always lower than unity. 
\vspace{4cm}

\section*{APPENDIX: The Negative Binomial Distribution as a Limit: Case $N=3$}\label{Apendix}
In this Appendix we show explicitly that, for $N=3$, the limit of eq. \ref{PN_de_n} when the three parameters are equal is eq. \ref{PN_de_n_BinNeg}. For simplicity in the notation, let us call $a_1=a$, $a_2=b$, $a_3=c$.
Eq. \ref{PN_de_n} for $N=3$ reads as follows:

\begin{eqnarray}  \label{P3_de_n}
\frac{P_3(n)}{K}=\frac{a^{n-1}}{(a-b)(a-c)}+\frac{b^{n-1}}{(b-a)(b-c)}+\frac{c^{n-1}}{(c-a)(c-b)} \nonumber \\  
\\   
K=(1-a)(1-b)(1-c) \nonumber   \\   \nonumber
\end{eqnarray}

To carry out the limit, we introduce new variables $x$ and $y$.

\begin{eqnarray}  \label{xy_def}
a=xc \nonumber \\  
\\ 
b=yc \nonumber  \\   \nonumber
\end{eqnarray}

The limit we seek will be implemented by tending $x$ and $y$ to 1.
Substituting the new variables into eq. \ref{P3_de_n}, the result is:

\begin{eqnarray} \label{P3_de_n_xy1}
\frac{c^{n-1}x^{n-1}}{c^2(x-y)(x-1)}+\frac{c^{n-1}y^{n-1}}{c^2(y-x)(y-1)}+\frac{c^{n-1}}{c^2(1-x)(1-y)}= \nonumber \\
\\
= c^{n-3}[\frac{x^{n-1}(y-1)-y^{n-1}(x-1)+(x-y)}{(x-y)(x-1)(y-1)}] \nonumber \\ \nonumber
\end{eqnarray}

Elaborating eq.\ref{P3_de_n_xy1} slightly, we obtain :

\begin{equation}  \label{P3_de_n_xy2}
\frac{c^{n-3}}{(x-y)(x-1)(y-1)}[y(x^{n-1}-1)-x(x^{n-2}-1)-y^{n-1}(x-1)]
\end{equation}

Henceforth it is convenient to use the following type of polynomials:

\begin{eqnarray}  \label{Polynomials}
P_n(x)=x^n+x^{n-1}+x^{n-2}+\dots+x+1 \nonumber \\  
P_n(1)=n+1\\ 
P_n(x,y)=x^n+x^{n-1}y+x^{n-2}y^2+\dots+xy^{n-1}+y^n  \nonumber \\  \nonumber
\end{eqnarray}

These polynomials fulfil the so-called cyclotomic property, namely

\begin{eqnarray}  \label{cyclotomic_property}
(x^n-1)=(x-1)P_{n-1}(x) \nonumber \\  
\\ 
(x^n-y^n)=(x-y)P_{n-1}(x,y) \nonumber   \\   \nonumber
\end{eqnarray}

So, dividing the second factor in eq. \ref{P3_de_n_xy2} by $(x-1)$ we obtain

\begin{eqnarray}  \label{P3_de_n_Pol1}
\frac{c^{n-3}}{(x-y)(y-1)}[yP_{n-2}(x)-xP_{n-3}(x)-y^{n-1}]= \nonumber \\
= \frac{c^{n-3}}{(x-y)(y-1)}[y(P_{n-3}(x)+x^{n-2})-xP_{n-3}(x)-y^{n-1}]= \\
= \frac{c^{n-3}}{(x-y)(y-1)}[(y-x)P_{n-3}(x)+y(x^{n-2}-y^{n-2})] \nonumber \\ \nonumber
\end{eqnarray}

Now we divide the second factor of eq. \ref{P3_de_n_Pol1} by $(x-y)$

\begin{eqnarray}  \label{P3_de_n_Pol2}
\frac{c^{n-3}}{(y-1)}[yP_{n-3}(x,y)-P_{n-3}(x)]= \nonumber \\
\frac{c^{n-3}}{(y-1)}[y(x^{n-3}+x^{n-4}y+x^{n-5}y^2+\dots+xy^{n-4}+y^{n-3})-(x^{n-3}+x^{n-4}+\dots+x+1)]= \\
\frac{c^{n-3}}{(y-1)}[(y-1)x^{n-3}+(y^2-1)x^{n-4}+\dots+(y^{n-4}-1)x^2+(y^{n-3}-1)x+(y^{n-2}-1)]= \nonumber \\ \nonumber
=c^{n-3}[x^{n-3}+x^{n-4}P_1(y)+x^{n-5}P_2(y)+\dots+x^2P_{n-5}(y)+xP_{n-4}(y)+P_{n-3}(y)] \nonumber \\ \nonumber
\end{eqnarray}

Returning to eq. \ref{P3_de_n}, using eq. \ref{P3_de_n_Pol2}, and performing the limit $x,y \rightarrow 1$,  we obtain:

\begin{eqnarray}  \label{limit}
\lim_{x,y\to 1}{P_3(n)}=(1-c)^3c^{n-3}[1+2+3+\dots+(n-3)+(n-2)]= \nonumber \\
\\
(1-c)^3c^{n-3}\frac{(n-1)(n-2)}{2} \nonumber \\ \nonumber
\end{eqnarray}

This formula coincides with eq. \ref{PN_de_n_BinNeg} when $N=3$
\begin{eqnarray}  \label{Demo}
P_3(n)=(1-c)^{3}c^{n-3}\binom{n-1}{2} \nonumber \\
\\
n = 3, 4, \dots, \infty  \nonumber   \\   \nonumber
\end{eqnarray}

\begin{acknowledgments}
This work was supported by the Spanish DGICYT (Project FIS2010-19773). AFP  would like to thank Jes\'us Asin, Jes\'us Bastero, Leandro Moral and Carmen Sanguesa who always help with a smile. 
\end{acknowledgments}

\bibliography{earthquakes}

\end{document}